# CAVDM: Cellular Automata Based Video Cloud Mining Framework for Information Retrieval

P.Kiran Sree[*1], Inampudi Ramesh Babu[2], SSSN Usha Devi N[3]

[1] Research Scholar, Dept of CSE, JNTU Hyderabad, India

[2] Professor, Dept of CSE, Acharya Nagarjuna University, India

[3] Dept of CSE, JNTU-K, India

[*1] profkiran@yahoo.com; [2] drirameshbabu@gmail.com; [3] usha.jntuk@gmail.com

*Abstract*

Cloud Mining technique can be applied to various documents. Acquisition and storage of video data is an easy task but retrieval of information from video data is a challenging task. So video Cloud Mining plays an important role in efficient video data management for information retrieval. This paper proposes a Cellular Automata based framework for video Cloud Mining to extract the information from video data. This includes developing the technique for shot detection then key frame analysis is considered to compare the frames of each shot to each others to define the relationship between shots. Cellular automata based hierarchical clustering technique is adopted to make a group of similar shots to detect the particular event on some requirement as per user demand.

**Introduction**

A wide range of possible applications that require Cloud Mining of video databases includes: news broadcasting, military, video, education and training, cultural heritage, advertising, web searching, crime prevention, geographical information system (GIS) etc. These applications has vast collection of images in the corresponding video databases and can be mined to discover new and valuable knowledge. Cloud Mining of video databases aims to automate such a knowledge discovery process.

To help user finds and retrieves relevant video effectively and to facilitate new and better ways of entertainment, advanced technologies must be developed for searching and mining the vast amount of videos currently available on the web. Although valuable information may be hiding behind the data, the overwhelming data volume makes it difficult for human beings to extract them without powerful tools. Video mining system that can automatically extract semantically meaningful information (knowledge) from video databases. While numerous papers have appeared on Cloud Mining, few deals with video Cloud Mining directly. In video databases, knowledge discovery deals with non-structured information, for this reason we need tools for discovering relationships between objects or segments within video components. In general video must be first preprocessed to improve their quality. Subsequently, these video files undergo various transformations and features extraction to generate the important features from the video databases. With the generated features, mining can be carried out using Cloud Mining techniques to discover significant patterns. These resulting patterns are then evaluated and interpreted in order to obtain the final application's knowledge. The problem of video Cloud Mining combines the area of content-based retrieval, image understanding, Cloud Mining, video representation and databases . Many applications maintain temporal and spatial features in their databases, which cannot be treated as any other attributes and need spatial attention. To be more particular instead of simply asking ourselves 'what' knowledge, it plays an important role and these can be trivially handled by spatial and temporal role and these can be trivially handled by spatial and temporal Cloud Mining techniques. Finally Cloud Mining of video databases is essentially a task of learning from video data can, in principle, being applied for Cloud Mining purposes. In general Cloud Mining algorithms are aimed at minimizing I/O operations of disk resident data, whereas conventional algorithms are more concerned about time and space complexities, accuracy and convergence.

It can be simply stated that he objective of video Cloud Mining is the organizing of video data for knowledge exploring or mining. Where the knowledge can be explained as special patterns (e.g., events, clusters, classification, etc.), which may be unknown before the processing. Many successful Cloud Mining techniques have been developed through academic research and industry hence, an intuitive solution for video Cloud Mining is to use these strategies on video data.

## Steps of Proposed Frame Work

Figure 1 depicts the frame work of video Cloud Mining consists of several steps, of which the primary step is segmentation of a video, because we can not apply the Cloud Mining directly to retrieve the information from video or detect the pattern from video data. So division of video into small segments is done basically to support relational database. Then further key frame analysis and clustering technique are involved to analysis the video data as well as retrieve the information.

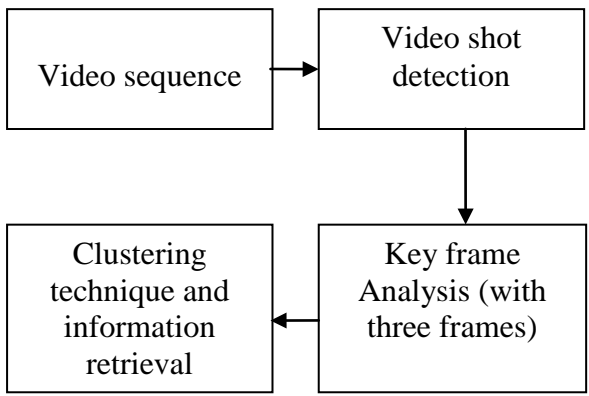

FIGURE 1 PROPOSED FRAME WORK OF VIDEO CLOUD MINING

*Shot Detection Technique*

The primary step of video Cloud Mining is to segment the video into small shots to represent into relational data format. Video segmentation, or shot change detection, involves identifying the frame(s) where a transition takes place from one shot to another. In cases where this change occurs between two frames, it is called a cut or a break. A shot is defined as a sequence of frames taken by a single camera with no major changes in the visual content.

Much technique has been developed to detect the shots. The most common approach to obtain boundaries of a shot is based on color information of the frame . In the recent studies, no matter which color space is of interest, main advantages of using color is its ease of implementation, descriptive characteristic both in spatial and temporal space and real-time applicability due to simplicity to obtain a feature vector, such as histogram. In this way reading the total no. of frames in the given video, after all divide the video into the shots on the basis of color observation of the frames. It is based on the application video, where a set is defined for the collection of the frames for each shot like $S_i = \{f^m, f^{m+1} \ldots f^n\}$.

Where *m* and *n* are the indices of the first and the last frames of the $i^{th}$ shot respectively. In the next section, we describe a method for compact representation of shots by selecting an appropriate number of key frames.

*Key Frames Analysis*

After detection of the shots from given medical video, key frame analysis is required to extract the features and define the relationships between the frames in shot. Choosing key frames of shots allows us to capture most of the content variations, due at least to camera motion, while at the same time excluding other key frames which may be redundant. The ideal method of selecting key frames would be to compare each frame to every other frame in the shot and select the frame with the least difference from other frames in terms of a given similarity measure. In this work (Figure-2) three frames of each shot consists of first, middle & last frame and it is represented as ($F_f$, $F_m$, $F_l$). Purpose for selecting three frames as key frames is to extract features correctly and more precisely.

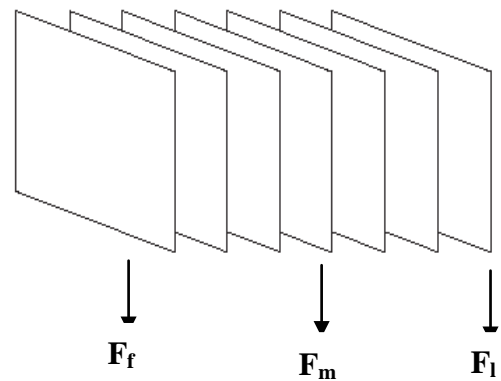

**FIGURE 2** THREE KEY FRAME IN A SHOT

## Clustering Techniques and Information Retrieval

Clustering similar shots into one unit will eliminate redundancy and produce a more concise video content summary. In unsupervised classification, the problem is to group a given collection of unlabeled video files into meaningful clusters according to the video content without a priori knowledge. Clustering algorithms can be categorized into partitioning methods, hierarchical methods, density-based methods, grid-based methods, and model-based methods.

### Definition of Neighborhood

For any point $P = (S_1, S_2, ..., S_K) \in P$, any text $d \in S$, a neighborhood point of P is defined as: let a text move from $S_i$ to $S_j$ then get a new partition $P'$. That is, if $d \in S_i$, and then the neighborhood of P on text d is:

$$N_d(P) = \{P\} \cup \{P' = (S'_1, S'_2, ..., S'_K) \mid S'_i = S_i - \{d\}, S'_j = S_j \cup \{d\}, S'_k = S_k, when, k \neq j, j = 1,2,...,K\}$$

Every neighborhood point of P correlates with a text. For any text $d \in S$, it has K different neighborhood points. The set of these neighborhood points is called neighborhood of P. The neighborhood of P includes $N \times K$ different neighborhood points.

### Search Strategy

In this paper, we select the steepest descent strategy. It can make an evaluation for every solution in a neighborhood of P, then choose one which can make the objective criterion function have maximal gain as a new solution. That is to say, it searches a candidate solution that can improve results furthest.

Suppose neighborhood of P is $Neighbour(P)$. The steepest descent strategy is to search a $P'$

$P' = argmax(E(P') - E(P) \mid P' \in Neighbour(P))$ in $Neighbour(P)$.

For any $p \in Neighbour(P)$, $E(P') \geq E(p)$ and $E(P') - E(P) > 0$

The text clustering algorithm based on Cellular automata based local search (TCLS) algorithm is composed of the following steps (one step):

1. For one clustering partition $P = (S_1, S_2, ..., S_K)$;

2. Suppose $max\Delta = 0$, $movedDoc = null$, $target = null$, ($movedDoc$ is the text that need to be moved, $target$ is the target class that $movedDoc$ moves to);

For every text $d \in S_i$ in S;

For all $j$, $j = 1, 2, ..., K \wedge j \neq i$,

calculate $\Delta_j E(P) \equiv E(P') - E(P)$

In P, let text d moves from $S_i$ to $S_j$, then get the $P'$;

Let $b = argmax\{\Delta_j E(P) > 0 \mid j \neq i\}$;

$max\Delta = max(max\Delta, b)$, $movedDoc = d$, $target = S_j$;

3. If $movedDoc \neq null$ (a best optimal solution has already been found), then let $d$ move from $S_i$ to target, and recalculate $D_i$ and $D_{target}$;

4. Return the partition $P'$.

The difference between Cellular automata based local search strategy and K-Means is:

Suppose $P = (S_1, S_2, ..., S_K)$,

$P' = (S'_1, S'_2, ..., S'_K) \in Neighbour(P)$ the target is to determine whether a text $d \in S_i$ should move to $S_j$.

For K-Means algorithm, it needs to test following inequality:

$$\Delta_k = d_0 \cdot (c_j - c_i) > 0 \qquad (3)$$

If $\Delta_k > 0$, then K-Means let $d_0$ move from class $S_i$ to $S_j$, else still let $d_0$ in class $S_i$.

Different from K-Means, Cellular automata based local search clustering algorithm calculate formula 4:

$$\Delta_{P'} E(P) \equiv E(P') - E(P) \qquad (4)$$

Derive formula 4 further, and according $S'_i = S_i - \{d_0\}$ and $S'_j = S_j \cup \{d_0\}$, then

$$\Delta_{P'} E(P) = \sum_{i=1}^{K} \sum_{d \in S'_i} d \cdot c'_i - \sum_{i=1}^{K} \sum_{d \in S_i} d \cdot c_i$$

$$= \left(\sum_{d \in S'_i} d \cdot c'_i - \sum_{d \in S_i} d \cdot c_i\right) + \left(\sum_{d \in S'_j} d \cdot c'_j - \sum_{d \in S_j} d \cdot c_j\right)$$

$$= \sum_{d \in S'_i} d \cdot (c'_i - c_i) + \sum_{d \in S'_j} d \cdot (c'_j - c_j) + d_0 \cdot (c_j - c_i)$$

So:

According Cauchy-Schwarz inequality,

$$\sum_{d \in S_i'} d \cdot c_i' \geq \sum_{d \in S_i'} d \cdot c_i ,$$

So: $\sum_{d \in S_i'} d \cdot (c_i' - c_i) \geq 0$

the same $\sum_{d \in S_j'} d \cdot (c_j' - c_j) \geq 0$

From upper two inequalities and formula 5:

From formula 6, even $\Delta_k \leq 0$, K-Means algorithm does not change adscription of $d_0$ in clustering results. The criterion function $\Delta_{p'}E(P)$ of LSKM is still right. Therefore, even if $E(P') > E(P)$, K-Means algorithm may miss partition $P'$.

*Clustering Algorithm Based on Cellular Automata Based Local Search*

Cellular Automata Based Local Search Based Clustering (LSC):

1. Give an initial clustering partition $P = (S_1, S_2, ..., S_K)$;

2. Run TCLS

3. If satisfy stop condition, then exit, else run step 2.

When the algorithm is running, the required space and time in every iterative are the same, the bottleneck of calculation is to calculate $\Delta_{P'}E(P) \equiv E(P') - E(P)$, that is for any $d \in S_i$,

Calculate: $E(S_i - \{d\}) - E(S_i)$ and

$$E(S_i - \{d\}) - E(S_i) = \|D_i - d\| - \|D_i\|$$
$$= \left(\sqrt{\|D_i\|^2 - 2\|D_i\|d \cdot c_i + 1} - \|D_i\|\right) \quad (7)$$

$$E(S_j \cup \{d\}) - E(S_j) = \|D_j + d\| - \|D_j\|$$
$$= \left(\sqrt{\|D_j\|^2 + 2\|D_j\|d \cdot c_j + 1} - \|D_j\|\right) \quad (8)$$

$E(S_j \cup \{d\}) - E(S_j)$

It can be noted that, in every iterative of K-Means, the $\|D_i\|$ and $d \cdot c_i$, $d \in S$, $i = 1,2,...,K$ are all need to be calculated.

$$\Delta_{P'}E(P) = \sum_{d \in S_i'} d \cdot (c_i' - c_i) + \sum_{d \in S_j'} d \cdot (c_j' - c_j) + \Delta_k \quad (5)$$

That is to say, the time, space and calculate complexity in every iterative of Cellular automata based local search or K-Means are almost the same.

For this work hierarchal clustering clustering is adopted to represent the cluster of the similar shots. An excellent survey of clustering techniques can be

$$\Delta_{p'}E(P) \geq \Delta_k \quad (6)$$

found. When clusters are generated to define the similar shots then events mining strategies is considered to evaluate the shots as per user demand. This technique is based on rules or visualization of the shots. After processing of the visualization on such ruleswhere particular information in form of shot is detected to achieve the information retrieval from video data.

## Conclusion and Future Work

This work proposes a cellular automata based video Cloud Mining frame work for information retrieval. It can be applied to many video applications to detect or retrieve the information. So this is the common process for video data miming. Future research work is to take a particular application and apply this frame work for video Cloud Mining. The main goal of this work is to develop the intelligence technique that can be used to support the video Cloud Mining process to retrieve information successfully and maintain the accuracy of the system.

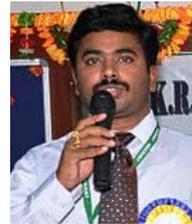

**P.KIRAN SREE** received his B.Tech in Computer Science & Engineering, from J.N.T.U and M.E in Computer Science & Engineering from Anna University. He is pursuing Ph.D in Computer Science from J.N.T.U, Hyderabad. His areas of interests include Cellular Automata, Parallel Algorithms, Artificial Intelligence, and Compiler Design. He was the reviewer for some of International Journals and IEEE Society Conferences on Artificial Intelligence & Image Processing. He is also listed in Marquis Who's Who in the World, 29th Edition (2012), USA. He is the recipient of Bharat Excellence Award from Dr GV Krishna Murthy, Former Election Commissioner of India. He is the Board of Studies member of Vikrama Simhapuri University, Nellore in Computer Science & Engineering stream. He is also track chair for some international conferences. He has published 30 technical papers both in international journals and conferences.

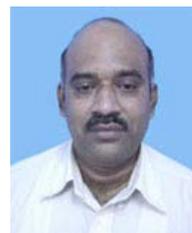

**INAMPUDI RAMESH BABU**, received his Ph.D in Computer Science from Acharya Nagarjuna University, M.E in Computer Engineering from Andhra University, B.E in Electronics & Communication Engg from University of Mysore. He is currently working as Head & Professor in the department of computer science, Nagarjuna University. Also he is the senate member of the same University from 2006. His areas of interest are image processing & its applications, and he is currently supervising 10 Ph.D students who are working in different areas of image processing. He is the senior member of IEEE, and has published 70 papers in international conferences and journals.